\documentclass[10pt]{emulateapj}
\usepackage{apjfonts}
\usepackage{psfig}
\usepackage{epsfig}
\usepackage{hyperref}

\hypersetup{
    pdfauthor   = {Nicole P. Vogt},
    pdftitle    = {An Online Tutor for Astronomy: The GEAS Self-Review Library},
    pdfsubject  = {Astronomy},
    pdfkeywords = {astronomy education research; astronomy education resources; college-level astronomy; distance education; STEM coursework},
    colorlinks  = true,
    linkcolor   = blue,
    filecolor   = magenta,      
    urlcolor    = cyan
}

\lefthead{Vogt \& Muise}
\righthead{An Online Tutor for Astronomy: The GEAS Self-Review Library}

\slugcomment{Accepted by Cogent Education 2015 March 27}

\begin{document}

\title{An Online Tutor for Astronomy: The GEAS Self-Review Library}

\author{Nicole P. Vogt$^1$ and Amy Smith Muise$^1$}

\affil{New Mexico State University, Department of Astronomy, \\
       P.O. Box 30001, Dept 4500, Las Cruces, NM 88003}
\email{nicole@nmsu.edu}

\begin{abstract}

We introduce an interactive online resource for use by students and college instructors in introductory astronomy courses. The General Education Astronomy Source (GEAS) online tutor guides students developing mastery of core astronomical concepts and mathematical applications of general astronomy material. It contains over 12,000 questions, with linked hints and solutions. Students who master the material quickly can advance through the topics, while under-prepared or hesitant students can focus on questions on a certain topic for as long as needed, with minimal repetition. Students receive individual accounts for study and course instructors are provided with overview tracking information, by time and by topic, for entire cohorts of students. Diagnostic tools support self-evaluation and close collaboration between instructor and student, even for distance learners. An initial usage study shows clear trends in performance which increase with study time, and indicates that distance learners using these materials perform as well as or better than a comparison cohort of on-campus astronomy students. We are actively seeking new collaborators to use this resource in astronomy courses and other educational venues. 

\end{abstract}


\section{Introduction}   

Students in introductory college science courses often arrive academically underprepared. In 2011, just 43\% of high school graduates met the American College Testing Program (\hyperlink{ACT2014}{ACT, 2014}) benchmark for mathematics, and only 37\% met the benchmark for science. High school preparation in mathematics is a significant predictor of success in introductory college science coursework (\hyperlink{Sadler2007}{Sadler \& Tai, 2007}), and it is sobering to realize that fewer than half of high school graduates currently meet the benchmark. Even more worrisome, of students interested in careers in education, only 37\% met the ACT math benchmark in 2011 (30\% for science) (\hyperlink{ACT2014}{ACT, 2014}). As 40\% of students taking general education science intend to become teachers (\hyperlink{Lawrenz2005}{Lawrenz, Huffman, \& Appeldoorn, 2005}), improvements can be leveraged to aid K--12 level educational efforts.

The last few years have seen tremendous growth in Internet access, with 95\% of American college students having home broadband access and 100\% having access through some source (\hyperlink{Pew2011}{Pew Research Center Internet \& American Life Project, 2011}). Online programs for homework, testing, and assessment in varied subject areas have been developing steadily in response (\hyperlink{Bennett2001}{Bennett, 2001}; \hyperlink{Koedinger2013}{Koedinger et al., 2013}), particularly within the sciences (cf., \hyperlink{Bonham2003}{Bonham, Deardorff, \& Beichner, 2003}; \hyperlink{Donovan2007}{Donovan \& Nakhleh, 2007}; \hyperlink{Kumar2005}{Kumar, 2005}). Opportunities for intensive, private, interactive study of key concepts can help students to develop skills without embarrassment, to gain knowledge, and to succeed. Online systems can provide repeated exposure to concepts, integration of mathematics with scientific material, and instantaneous feedback (\hyperlink{Nguyen2006}{Nguyen Hsieh, \& Allen, 2006}), as well as increasing access for students at a distance and/or asynchronously. 

There are several rich content resources for stand-alone astronomy questions, including Green's hundreds of ConcepTests (\hyperlink{Green2002}{Green, 2002}) and the thousand questions collected by the Collaboration of Astronomy Teaching Scholars (CATS; \hyperlink{Wallace2012}{Wallace et al. 2012}). Online tutors for astronomy are offered through several publishing houses, where these resources are typically bundled with textbooks, and students pay a sizable fee for access for a single semester (\hyperlink{Barmby2010}{Barmby, 2010}). Our self-review library, an interactive online tutor that works with students to develop mastery of core astronomical concepts and mathematical applications, can be distinguished both in terms of its scope (containing 12,000+ questions, providing detailed student diagnostics and individual, adaptive feedback), accessibility (runs in any desktop or mobile web browser) and affordability (free). Pilot usage studies have been conducted at New Mexico and California colleges, and our resources are now being offered nationwide.

\section{Structure of the GEAS Self-Review Library}

The General Education Astronomy Source (\href{http://astronomy.nmsu.edu/geas/}{GEAS}) self--eview library is an intelligent online tutor built around 26 modules, each based on a 75-minute lecture. The modules follow a common pattern for a semester-long course in general astronomy. Each module is presented as a series of web pages (200 in total), augmented with slide-by-slide audio lectures and reproductions of white board diagrams from conventional classroom lectures. A printed textbook is no longer required in our classroom. We have found (as have others, cf. \hyperlink{NAS2007}{National Academy of Sciences, 2007}; \hyperlink{Stavrianeas2008}{Stavrianeas, Stewart, \& Harmer, 2008}) that the majority of the students perform well with alternative, no-cost, resources. A set of 440 Think-Pair-Share questions (\hyperlink{Lyman1981}{Lyman, 1981}) have also been created, developed for the modules in the form popularized via National Aeronautics and Space Administration (NASA) Center for Astronomy Education (CAE) workshops (cf. \hyperlink{Forestell2008}{Forestell, et al. 2008}).

Material is presented at a level accessible for non-science majors, with a balance between qualitative and scaling arguments and numerically motivated ideas. The basic mathematics needed for the material (algebra, roots, exponents, and scientific notation) is presented early on and reinforced throughout the sequence. The art of visualization is presented as an important problem-solving strategy and emphasized for topics for which it is most useful (e.g., understanding the phases of the Moon, eclipses, and orbital mechanics).

\begin{figure*} [htbp]
  \begin{center}\epsfig{file=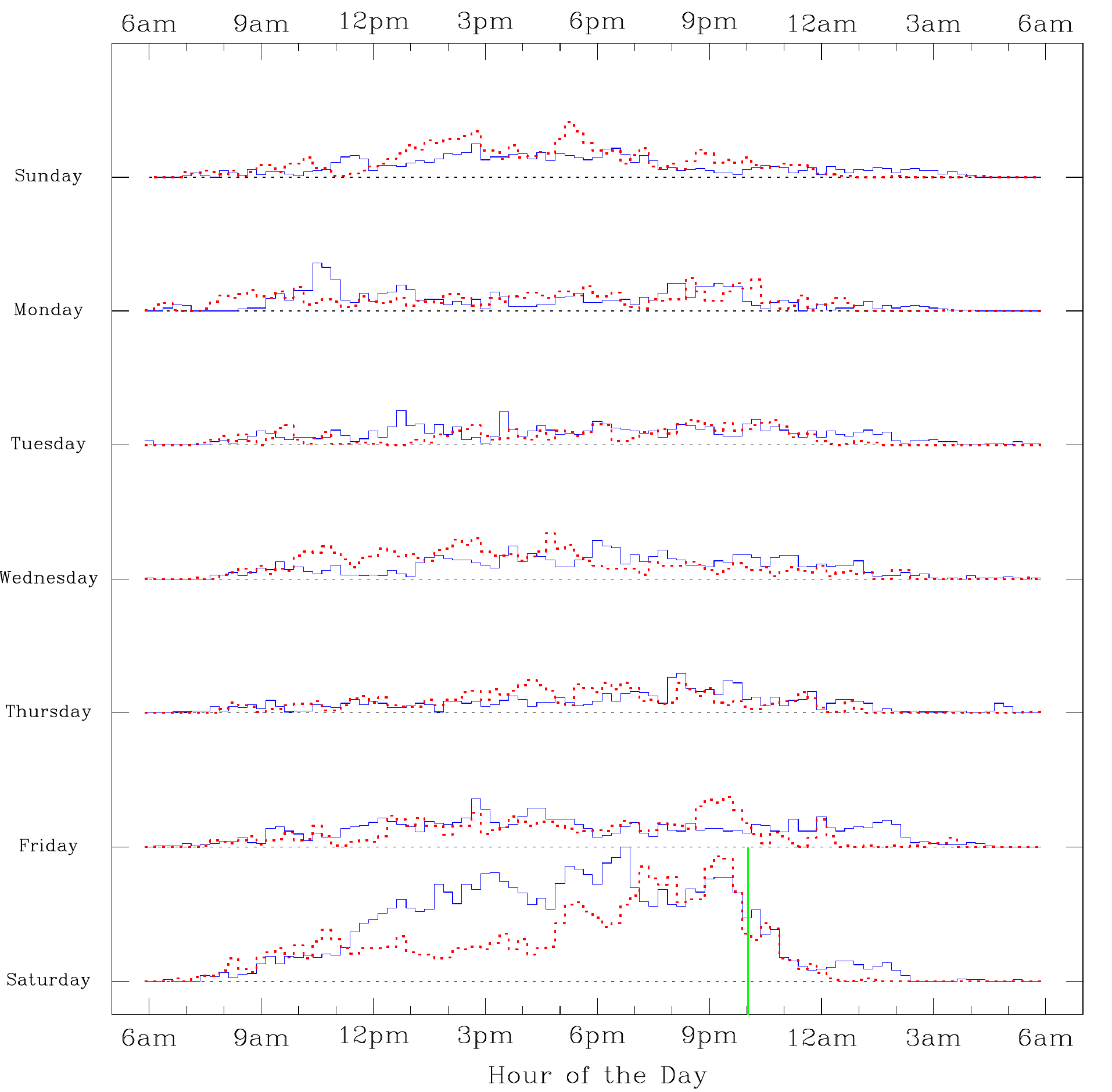, width=5.0in}\end{center}
  \caption[Figure 1] 
{The weekly pattern of self-review library usage for distance learning students over a semester. The red dotted line shows the distribution of 7,160 questions completed by 22 students who ranked the library as a good study tool, better than average, and the blue solid line shows the distribution of 10,210 questions completed by 27 students who ranked it as one of the best tools ever (5 students who ranked it as typical are not shown). Histograms are scaled so that equivalent amounts of work per student within each group would produce histograms of the same height. The first group answered 102 questions on average each week (towards a goal of 100), while those who found the tool most useful completed an additional 16 questions per week. Both groups had similarly low study times during the week and focused their efforts during the three hours before the weekly 10 pm deadline on Saturdays (marked by a vertical green line). The second group studied significantly more from 11 am to 7 pm on Saturdays, rather than adding in additional study periods on other days of the week. The self-review library offers students immediate, detailed feedback outside of business hours and the flexibility of being able to study at all hours of the day or night.}
  \label{fig01}
\end{figure*}

\subsection{Pattern of Usage each Week}

Figure 1 shows the distribution of self-review library usage, identifying study patterns each week for a cohort of distance learners. While most studying took place between noon and midnight, there is a tail of activity extending out past 3 a.m., and another picking up at 6 a.m. (e.g., a night watchman studied primarily in twenty-minute blocks spread out between midnight and eight am, in between making his rounds.) Individuals work in blocks of time of average length 22 minutes; the broad features shown thus represent the overlapping efforts of multiple students. Efforts are concentrated before weekly deadlines. 

Students were divided into two groups based on their ratings for the tutor after a semester of usage. They responded positively overall to the tool, with 9\% finding it acceptable, 41\% classifying it as a better than average study tool and another 50\% rating it as one of the best study tools they had ever seen. Figure 1 contrasts the study patterns of these last two groups. We find that the most enthusiastic users completed 16\% more problems while studying, and tended to study significantly more in the afternoons leading up to weekly deadlines.

\begin{table*} [htbp]
  \caption{Self-Review Library Question Distribution}
  \begin{center}
  \begin{tabular} {l c r r r r} 
  \hline
  \hline
  Module & $\mu \pm \sigma^1$ & AT$^2$ & MC$^3$ & NV$^4$ & Total \\
  \hline 
  1. The Contents of the Universe & 80.2 $\pm$ 0.5 & 66 & 96 & 31 & 127 \\ 
  2. Scientific Notation & 68.9 $\pm$ 0.6 & 45 & 112 & 626 & 738 \\ 
  3. Timescales in the Universe & 70.9 $\pm$ 0.7 & 53 &  91 &  47 & 138 \\ 
  4. The Phases of the Moon & 58.4 $\pm$ 0.7 & 57 & 597 &  65 &  662 \\ 
  5. The Seasons on Earth & 54.9 $\pm$ 0.7 & 53 & 157 & 164 & 321 \\ 
  6. The Origin of the Moon   & 71.5 $\pm$ 0.7 & 56 &  81 & 220 & 301 \\ 
  7. The Celestial Sphere   & 67.9 $\pm$ 0.7 & 59 & 173 &  99 & 272 \\ 
  8. Planetary Orbits   & 65.9 $\pm$ 0.7 & 46 &  99 &  47 & 146 \\ 
  9. The Scientific Method   & 68.2 $\pm$ 0.7 & 75 & 732 & 2,534 & 3,266 \\ 
  10. Geocentric and Heliocentric Models   & 61.2 $\pm$ 0.8 & 63 & 116 & 33 & 149 \\ 
  11. The Formation of the Planets   & 72.7 $\pm$ 0.7 & 48 &  83 & 31 & 114 \\ 
  12. The Terrestrial Planets   & 67.4 $\pm$ 0.8 & 69 &  85 &  41 & 126 \\ 
  13. The Jovian Planets   & 67.4 $\pm$ 0.8 & 50 &  84 &  48 & 132 \\ 
  14. Waves and Light   & 72.2 $\pm$ 0.7 & 55 & 1,499 & 1.004 & 2,503 \\ 
  15. Atomic Structure   & 62.5 $\pm$ 0.8 & 46 &  392 &  40 & 432 \\ 
  16. Absorption and Emission   & 64.7 $\pm$ 0.8 & 60 & 153 &  55 & 208 \\ 
  17. Stellar Temperatures   & 69.0 $\pm$ 0.8 & 60 & 150 &  54 & 204 \\ 
  18. Nuclear Reactions   & 57.3 $\pm$ 0.9 & 53 & 144 &  47 & 191 \\ 
  19. Binary Stars   & 60.0 $\pm$ 0.9 & 51 & 147 & 220 & 367 \\ 
  20. The Hertzsprung-Russell Diagram   & 65.9 $\pm$ 0.8 & 48 & 180 & 860 & 1,040 \\ 
  21. White Dwarfs   & 59.5 $\pm$ 0.8 & 50 &  99 &  30 & 129 \\ 
  22. Neutron Stars   & 66.2 $\pm$ 0.8 & 48 &  93 &  48 & 140 \\ 
  23. Black Holes   & 61.8 $\pm$ 0.8 & 46 &  81 &  30 & 111 \\ 
  24. The Milky Way   & 62.1 $\pm$ 0.9 & 46 & 200 &   87& 287 \\ 
  25. The Expansion of the Universe   & 74.5 $\pm$ 0.8 & 61 &  80 &  55 & 135 \\ 
  26. A Universe of Galaxies   & 67.0 $\pm$ 1.3 & 69 & 111 &  55 & 166 \\ 
\hline
  Average Number of Questions per Module   & & 55 & 224 & 253 & 477 \\ 
  Total Number of Questions   & & 1,434 & 5,835 & 6,571 & 12,406 \\ 
 \hline \\
\multicolumn{6}{l}{$^1 ~\mu \pm \sigma$: Mean and standard deviation of the mean for student answers} \\
\multicolumn{6}{l}{$^2$ AT: Archetypes of independent questions, from which entire families are formed} \\
\multicolumn{6}{l}{$^3$ MC: Multiple choice, ranking, or figure evaluation questions} \\
\multicolumn{6}{l}{$^4$ NV: Numerical value questions}
  \end{tabular}
  \end{center}
  \label{tab01}
\end{table*}

\subsection{Topical (Lecture) Modules}

Table 1 lists the 26 topic modules and the number of self-review questions of various types for each module. The sequence begins with a short set of three modules (1--3) that present an overview of the astronomical sequence and review the math skills necessary for the entire sequence. Questions within these modules are designed to be more straightforward and are written at a simpler level than the average, to give students a chance to become familiar with and to adjust to using the interface. The next two modules (4--5) introduce the concept of developing a model (through the scientific method) that enables us to predict the future behavior of physical objects. Modules 6--8 and 10--13 focus on the history and behavior of objects within the solar system. There is a continued emphasis on visualization when appropriate, as well as presentation of new astronomical knowledge. Module 9 focuses on the scientific method. It also contains a review of additional mathematical applications, including linear fits and the concepts of a histogram, a normal distribution, a mean value, and a standard deviation, useful for laboratory experiments (\hyperlink{Vogt2013}{Vogt, Cook, \& Muise, 2013}). These first 13 modules comprise the first half of the sequence and fall loosely under the umbrella of ``solar system'' material.

The second set of modules covers stellar and extragalactic astronomy topics. Modules 14--15 introduce the concepts of light as a wave with an associated frequency and energy and the modern model of the atom. In modules 16--18 the absorption and emission of photons within the atom is connected to macroscopic observables such as stellar spectra for stars of various masses and temperatures. Module 19 delves into Doppler shifts and binary star systems, and module 20 combines the ideas developed in 16--19 to motivate the Hertzsprung-Russell Diagram of stellar evolution. Modules 21-23 describe the life cycles of stars of various masses. The final three modules (24--26) introduce ideas about galaxies and the cosmological structure and expansion of the Universe.

\subsection{Self-Review Questions}

Table 1 lists the number of self-review questions for each module, 12,406 in total, divided into several groups of interest. The variations in the mean values of student attempts by lecture reflect variations in difficulty between topics. The two modules (4--5) with the lowest means develop the ability to visualize three-dimensional movements of solar system objects such as the Earth, Moon, and Sun system, and the next lowest (Module 18) introduces the mathematically challenging concept of radioactive decay rates. 

There are groups of similar questions within each module, all derived from one template/archetype problem.  This is particularly true for the numerical and spatial-visualization problems which students find most challenging. The column labeled ``AT'' defines the number of archetype questions that are independent from one another (having different content, and not containing large clues to each other's answers). Questions labeled as ``MC'' are multiple-choice questions -- a category which contains both traditional multiple choice questions, ranking tests, and figure analysis exercises -- while those labeled ``NV'' are numerical value questions. For these questions, the student enters a number rather than selecting one of several answers, the value of which must lie within a certain range to be considered correct. The final column lists the total number of questions within each module. There are on average 55 archetypes per module, and 477 questions based on these archetypes split fairly evenly between the MC and NV modes. Each individual question contains three components: the question proper, a tailored hint that provides guidance but does not reveal the answer, and a fully worked solution to the problem. 

Questions are designed for each module to give experience in a variety of problem-solving modes, and with an emphasis on quantification and on extrapolation, drawing on physical evidence and theory to deduce logical properties and patterns found in the physical universe. Question types include the following:

\newcounter{bean}
\begin{list} {(\Alph{bean})} {\usecounter{bean} \topsep -0.1in \itemsep -0.05in}
\item Extrapolation: How can our knowledge of the biodiversity found
  on Earth enable us to estimate the probability of finding life in
  the oceans of Europa? [Modules 12 and 13]
\item Scaling: If different galaxy components have different spectral energy distributions (emitted flux as a function of wavelength), how should optical and infrared images of the bulge or disk of a spiral galaxy like the Milky Way differ from each other? [Module 26]
\item Visualization: Given a physical model of the Sun-Earth-Moon system, what is the observed phase of the Moon at a certain position in the sky at a certain time of day? [Module 4]
\item Figure analysis: What is the frequency of a displayed light wave? [Module 14]
\item Computation: If the Sun burns its reservoir of hydrogen at a certain rate, how long can it exist in the hydrogen-burning phase? [Module 17]
\item Algebra: If the Hertzsprung-Russell diagram shows us the specific relationship between the luminosity, temperature, and radius of a star, how can one calculate one quantity from the other two variables? [Module 17]
\end{list}
\vspace{0.15truein}

Students interact with the self-review library by requesting a five-question quiz. Each question is presented with two aids, a link to a tailored hint, and a link to the appropriate lecture slide from the parent module. Students can thus refresh the connection between the question and general topics, or receive guidance on how to set up or think about the problem. After answering all five questions, students submit their work. They are presented with a solution set two seconds later, one which shows both the correct answers and the submitted answers for comparison. 

Students may choose to review a single module at a time or to study a sequence of modules. The selection process prioritizes all modules within a sequence equally and all archetypes within a selected module equally, drawing from any archetype no more than once per quiz. All chosen modules thus have an equal probability of appearing on a quiz, and archetypes used to generate large sets of questions are not over-represented. Archetypes are not re-shown to a given student until all other archetypes within the parent module have been seen at least once, so that students encounter a variety of questions covering all of the concepts within a module on their first pass through new materials. Questions within a single archetype are also not re-used until all versions have been used at least once, so that when archetypes re-appear they contain new specific content. Students can thus review new concepts such as the relationship between the energy, wavelength, and color of an absorbed or emitted photon involved in an atomic transition for as long as they wish without ever repeating exact questions. 

Students may also override the default selection process and enter challenge mode, where the archetypes which have been the most challenging for them initially are prioritized in populating quizzes. This is particularly useful to students once they think that they have mastered a new module, as they can then focus on the types of questions which gave them the most difficulty and prove themselves competent. Students typically study large sequences of modules for exams, and find the challenge mode to be useful then for focused review of targeted topics.

\subsection{Self-Review Solutions}

Each solution set begins with a header noting the total score and a list of the questions in abbreviated form indicating quickly which ones were answered correctly. A complete solution is then presented, including a discussion of how decisions were made along the way. Equations are embedded into the discussion as small PNG-format images. In the case of mathematical questions, a fully worked solution is presented with the exact numerical values used in the particular problem (not using variables to illustrate a general solution), so that students may check their work at every stage of the problem.

\subsubsection{An Instructor Immediacy and Affective Behavior Algorithm}

There is evidence that emotional state plays a role in learning (cf., \hyperlink{Craig2004}{Craig et al. 2004}). Studies have also shown that instructor immediacy, utilizing verbal and non-verbal communication techniques which act to reduce the perceived distance (social and psychological) between instructor and student, has a positive effect on instructional outcomes and results in increased cognitive and affective learning (\hyperlink{Andersen1979}{Andersen, 1979, pp. 543-559}; \hyperlink{Gorham1988}{Gorham, 1988}; \hyperlink{Witt2004b}{Witt, Wheeless, \& Allen, 2004}). This type of analysis is being extended from evaluating in-class interactions to examine online interactions as well (\hyperlink{Arbaugh2001}{Arbaugh, 2001}; \hyperlink{Rourke2001}{Rourke et al., 2001}; \hyperlink{Witt2004a}{Witt, 2004}). An adaptive element, in which tutors provide targeted materials and feedback in response to individual student activity, is also acknowledged to be extremely effective (cf., \hyperlink{Murray1999}{Murray, 1999}). 

The self-review library is designed to support student efforts to study privately yet interactively. Students, especially women, often mention this aspect positively -- they appreciate that they can study a taxing concept for as long as it takes to learn it, without revealing, as they would if studying with peers, that it was difficult for them. However, because the system does grade the submitted material, users tend to relate to it as a judge rather as a mentor or guide. Initial usage results suggest that students who view the library more as a personal guide and supporter and less as a taskmaster will engage more freely, and more often, with difficult topics, in accordance with recent studies in affective learning (\hyperlink{Burleson2006}{Burleson, 2006}).

Human facilities studies (\hyperlink{Fogg1997}{Fogg \& Nash, 1997}; \hyperlink{Nass2004}{Nass, 2004}) have shown that humans frequently interact with computers automatically, applying and responding to social cues as they would with people, given even fairly minimal encouragement from a device that they know intellectually is not a social entity. People will be more likely to do a favor for a computer that has helped them (displaying reciprocity), and will treat a computer more kindly (with praise and greater attention) if it, or even another computer with a similar appearance, has recently provided them with helpful information.

Two strong, basic cues of social support amongst humans are naming and empathy. A pilot algorithm has thus been launched examining how the interface can encourage students' usage of the self-review library by addressing them by name and giving adaptive,  positive feedback -- rewarding correct work and appropriate study patterns (such as spending appropriate amounts of time working on problems) -- and by offering sympathetic prompts in response to poor scores. 

Each solution set contains feedback provided by an adaptive routine that evaluates the accuracy of the student's answers, the amount of time spent per question, and the number of questions answered to date on a particular topic. The goal of this algorithm is to motivate students to succeed, encourage them to improve their study habits by providing appropriate feedback, and to present a more social persona for the self-review library through interaction.

\subsection{Programmatic Implementation}

The self-review library is implemented in dynamic HTML derived from an ASCII database on an Apache HTTP server. The HTML feature set is as conservative as possible, to allow for usage on both recent and outmoded versions of a range of web browsers. Every effort has been made to keep data file sizes to a minimum to increase accessibility. User names and access codes are stored in an encrypted format, and the back-end database controls have been written in Perl and C, optimized so that user requests take less than two seconds to fulfill.

\subsection{Student Portal and Interface}

Students cohorts enter the library through dedicated portals. They can select a single module or a range of modules to review, with default options updated every week to match their class schedule of topics. The article appendix contains examples of a quiz, a question hint, and a solution set.

When students are confident that they have mastered certain modules through review, they can opt to take one of 16 weekly ``formal'' quizzes. Most weekly quizzes cover two modules at a time, and at the sequence half-point and end-point there are summary versions that cover either half or all of the modules together. The key difference between review and weekly quizzes is that review quizzes (in effect, homework) contain links back into the supporting web pages and guided hints for each question, while weekly quizzes contain no such aids. Each weekly quiz is available for a full three-week window of time, to allow students to work ahead or behind by a week as necessary. 

At New Mexico State University (NMSU), homework scores are constructed from the review quizzes; students receive the higher of (a) their average review quiz score, or (b) the average number of review questions completed per week (up to 100). A student who takes 20 five-question quizzes per week on average is thus guaranteed a homework score of 100\%, while one who takes 16 quizzes per week is guaranteed a homework score of 80\% (and receives that score or their average review quiz score if it is higher than 80\%). Our policy encourages students to complete 100 questions per week, a goal achievable with several hours of study per week, which exposes them to a reasonable amount of content review and practice in problem-solving. The average score for the 16 weekly quizzes is used to construct an independent course quiz grade, with no adjustment for effort.

\begin{figure*} [htbp]
  \begin{center}\epsfig{file=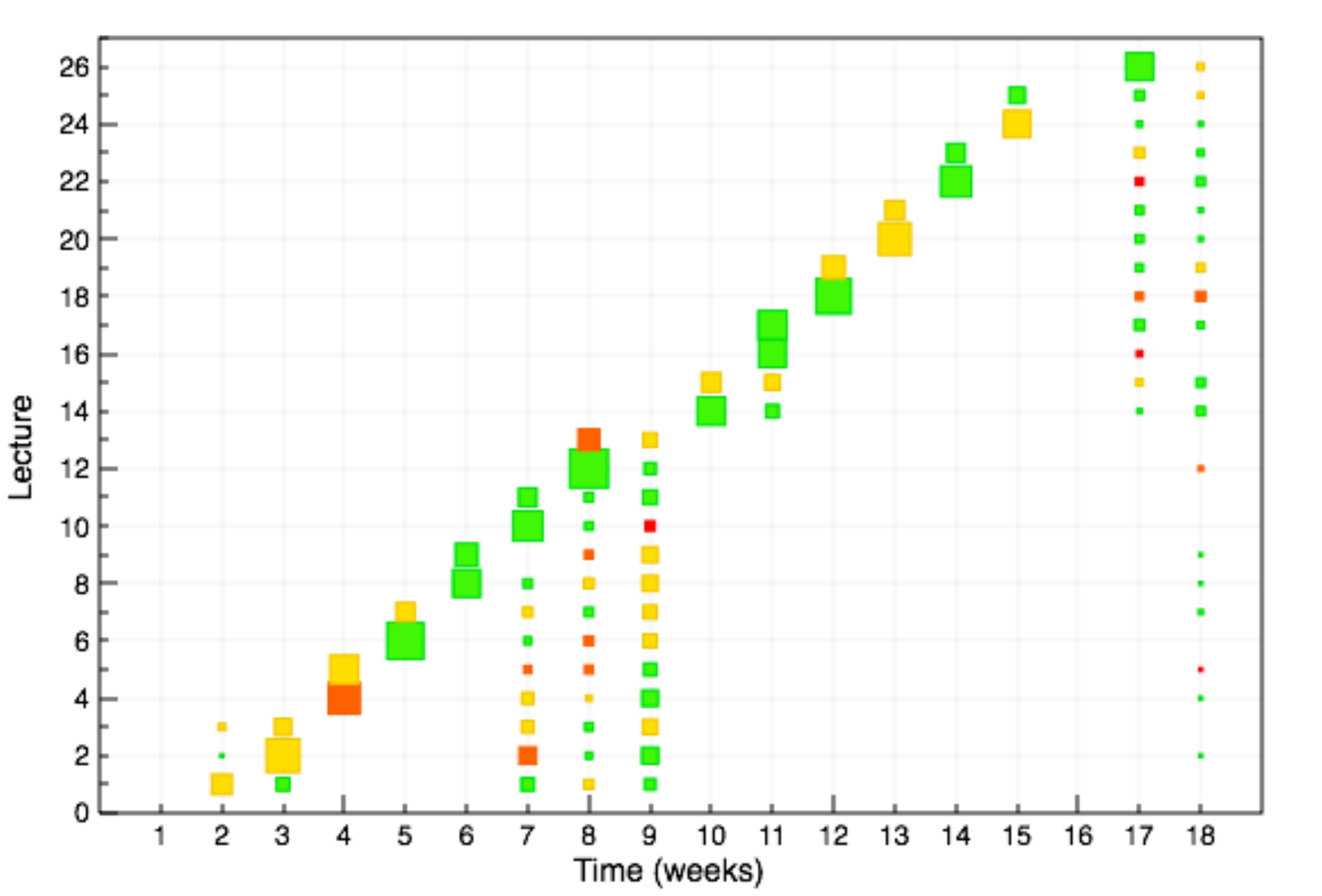, width=5.0in, clip=true}\end{center}
  \begin{center}\epsfig{file=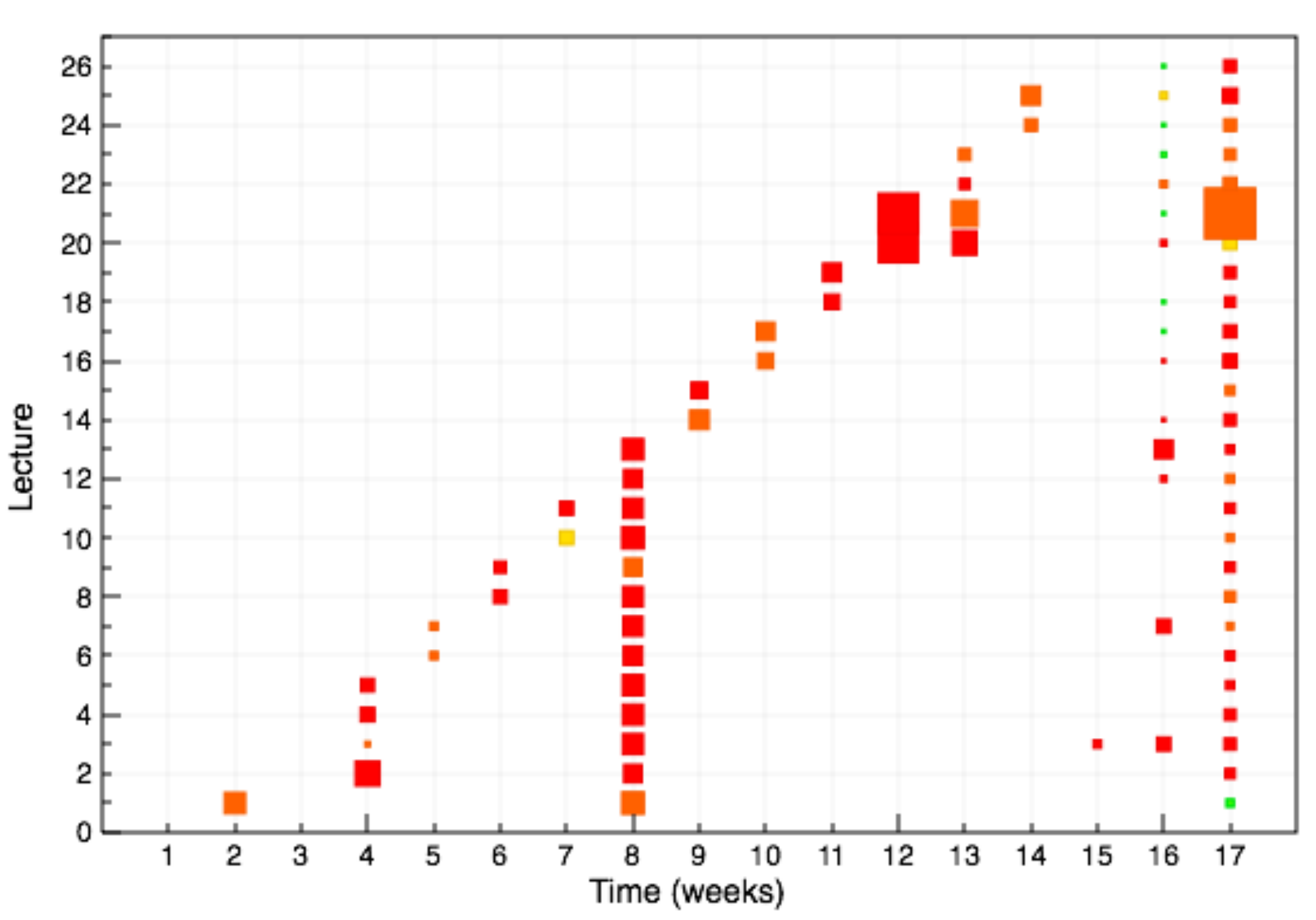, width=5.0in, clip=true}\end{center}
  \caption[Figure 2] 
{Status indicators for two representative students, taken from progress reports which can be viewed at any time by students and their instructors. Each figure shows the number of questions answered within each lecture module over time, with the number of questions attempted each week scaling with the area of each box. Results are color-coded to indicate success rates above 80\% (green), between 65\% and 80\% (yellow), between 50\% and 65\% (orange), and below 50\% (red). The top panel shows a record for a successful student, while the bottom panel shows the performance of a student having grave difficulties.}
  \label{fig02}
\end{figure*}

\subsection{Progress Reports}

Students have access to progress reports which can be generated on the fly at any time. They show the number of quizzes completed and present average scores, including a breakdown into multiple choice and numerical value components. (Scores for numerical problems, which require students to perform a calculation and enter a numerical answer, are typically lower than those for other questions.) Two figures show the average scores and number of questions attempted by week and by lecture against clear target goals, to identify key modules for additional study. A third overview figure contains multiple indicators of progress, as shown in Figure~2. A table of all weekly quiz scores is also provided so that students can monitor their progress and verify which weekly quizzes have been completed.

\begin{figure*} [htbp]
  \begin{center}\epsfig{file=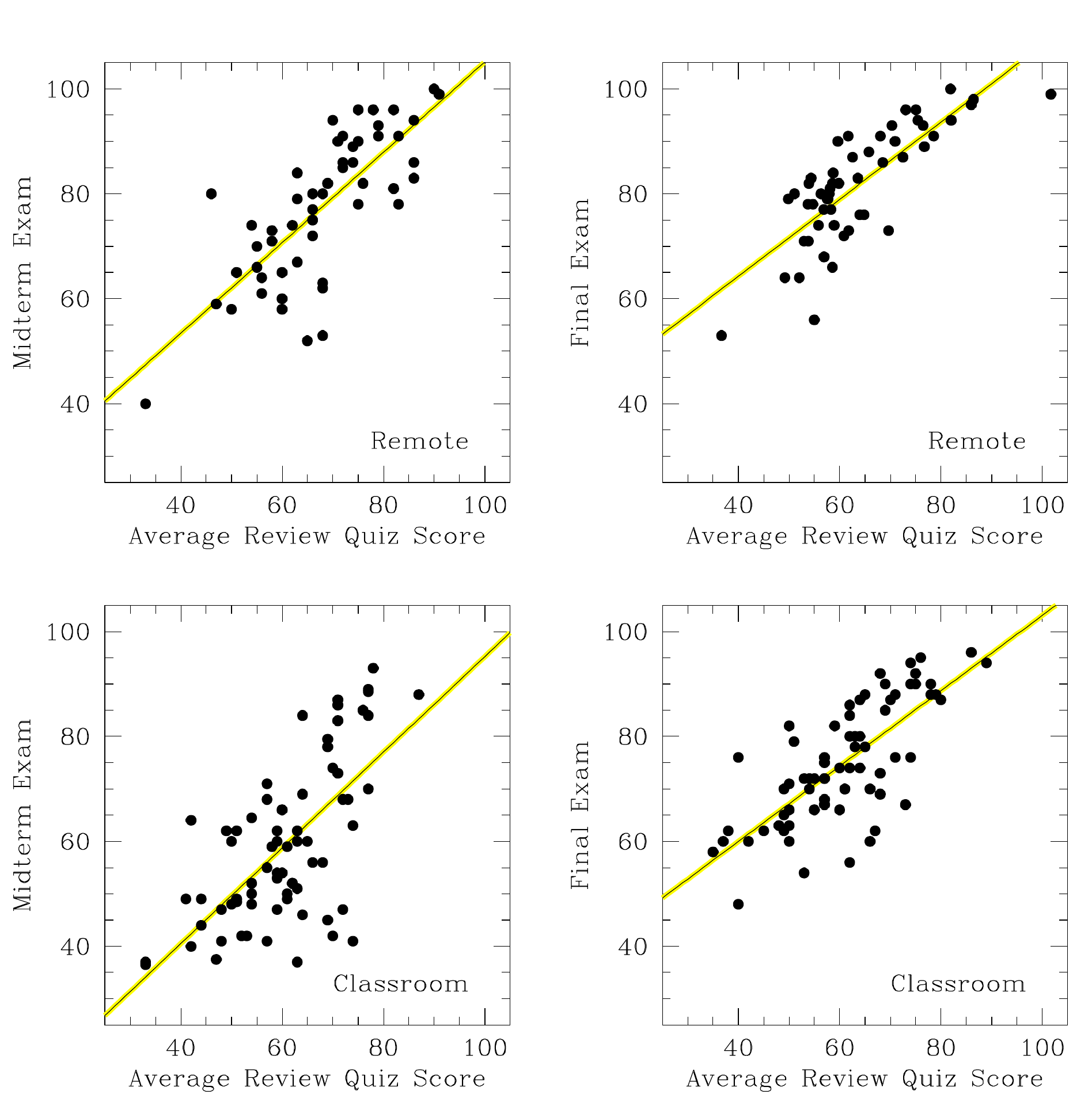, width=5.0in}\end{center}
  \caption[Figure 3] 
{Relationship between exam scores and efforts within the self-review library, comparing the results of tool usage to traditional indicators of performance for distance learners (top panels) and their in-class peers (bottom panels). The panels show the relationship of midterm and final exam scores to average review quiz (homework) scores at exam time. Linear fits are consistent with the data distribution, and correlation coefficients range between 60\% and 80\%. Note that the locus of points is shifted to higher exam and homework scores for the distance learners.}
  \label{fig03}
\end{figure*}
 
\subsection{Feedback Reports}

Students also have an opportunity to submit feedback after each quiz. This provides a mechanism for flagging any errors in question implementation in real time, and also allows students to work directly with us if they feel that there are issues with questions or wish to simply give us their opinions. This tool is particularly helpful for students taking classes outside NMSU (at other institutions), as it allows them to interact directly with the self-review developer (NPV) rather than having to funnel their feedback through their own instructors. Small extra credit ``bounties'' are offered for finding significant errors, but more importantly, we benefit from receiving student impressions of questions in real time -- very helpful when adjusting scientific terms or figure points of view to be most accessible.

\subsection{Instructor Resources}

Instructors are provided with information for entire cohorts and for individual students. Summary HTML-format tables present average scores per module topic for each student (and group averages) and weekly quiz scores, as well as breaking down results into multiple choice and numerical value questions, and evaluating review (homework) and the weekly quiz averages separately. 

Instructors can click on the name of any student from within the summary tables to review individual work. The interface will present for each student the materials within the student progress reports (described above), plus several enhancements. Tables break down the average scores and number of questions by module topic and by week, and sort most-missed question archetypes by lecture, ranked by student success so that the hardest questions for each student can be easily pulled up for review. 

Figure 2 shows two instructor overview figures of student progress, also available for students to view. The first student record (top panel) represents successful learners, with a healthy mix of green and yellow boxes overall and few orange indicators of difficulty. Filled-in columns represent study of topics-to-date for midterm and final exams; note that the colors for modules 4 and 13 (originally orange) shift to yellows and greens when reviewed before the midterm exam. The second student record (bottom panel) represents students having serious difficulty, as evidenced on first glance by the sea of red points. Boxes are smaller, indicating fewer study questions have been attempted, and even when additional efforts are made (see modules 20 and 21) average scores do not rise significantly. The columns of red boxes during exam study periods indicate no additional  success when reviewing materials. Another common signature of difficulty (not shown on this record) is blank weeks where no work is done, followed by bouts of catch-up activity in which too many topics are covered at once for deep learning to occur. These figures can help instructors to identify potential problems of these types quickly, so as to intervene and provide additional guidance on productive studying habits. 

Instructors are also provided with tables listing all quizzes and all quiz questions completed by each student, sorted by module or by date and divided into review mode, weekly mode, or showing all work. The date and time are shown for each quiz, as is the amount of time spent. Each entry links to a complete quiz recreation, showing student answers and correct answers and containing the original links to hints and lecture web pages. These quiz recreations can be very useful when meeting with a student one-on-one. They enable the instructor to discuss the amount of studying being done and the timing of the work, and allow instructors and students to review quizzes question by question and discuss appropriate strategies. They enable instructors to show explicitly how the information in a hint can be used to answer a question, or to identify patterns of mistakes shown in student answers (such as repeatedly multiplying rather than dividing in a particular type of units conversion). 

Instructor reports are compiled daily, and securely distributed as encrypted, compressed archives that expand into a directory of HTML files conveniently navigated with any web browser.

\subsection{Additional Self-Review Library Data Analysis Options}

A wealth of information is logged, allowing detailed analyses to be conducted on a variety of factors. Results are stored in an open format and can be analyzed with any tool (e.g., a spreadsheet or scripts). One can study the amount of time spent solving problems and reading solution sets, count the number of questions answered per topic or per unit time, plot the amount of pre-quiz practice done versus scores on weekly quizzes, compare scores for self-review and exam questions by topic, and track progress over time on a topic for individuals or for a cohort. One can also ``invert the analysis'' and evaluate questions, comparing success rates for different questions (or sets) or for a question when answered at different times (e.g., on start, or after studying a particular topic for an hour).

\section{A COMPARISON OF USAGE OUTCOMES (REMOTE AND CLASSROOM COHORTS)}

It is reasonable to inquire how the use of online tutors affects learning. Previous studies (\hyperlink{Aleven2002}{Aleven \& Koedinger, 2002}) have established a hierarchy of resources, ranking self-study as least effective followed by traditional classroom instruction 0.75 standard deviations ($\sigma$) above (\hyperlink{Corbett2001}{Corbett, 2001}). Typical one-on-one human tutors rank 0.4$\sigma$ higher (\hyperlink{Cohen1982}{Cohen, Kulik \& Kulik, 1982}; \hyperlink{Graesser1995}{Graesser, Person \& Magliano, 1995}), with intelligent computer-based tutors another 0.6$\sigma$ higher (\hyperlink{Anderson1995}{Anderson et al., 1995}; \hyperlink{Koedinger1997}{Koedinger et al., 1997}), topped by the best human one-on-one tutors a full standard deviation above computerized systems based on  meta-analyses of student performance (\hyperlink{Bloom1984}{Bloom, 1984}; \hyperlink{Kulik1991}{Kulik \& Kulik, 1991}). There is also some evidence that computer-based systems allow students to master new material faster (\hyperlink{Kulik1991}{Kulik \& Kulik, 1991}). With the use of such tools we seek not to replace human interaction, however, but rather to augment it, supporting tailored instruction in situations where student-teacher ratios are highest and where students and teachers are isolated from each other and may work asynchronously. 

The self-review library utilizes several techniques deemed most efficient for learning. First, it employs distributed, rather than massed, practice of problem types, and constantly interleaves practice of multiple topics. Students typically study in multiple 20-minute sessions spread throughout each week of the semester, and each review session is focused around two lecture modules containing 100 question archetypes and covering six key concepts for the week. Rather than reading full textbook chapters and then switching over to solving problems in blocks, our materials are interwoven so that on a question-by-question basis students can consider a single problem, review a single lecture slide, listen to supporting audio describing applications of new concepts, and process a hint that guides them in how to set-up the problem successfully. This helps students to read new material in an informed manner, constantly assessing what can be derived from it to apply to a range of relevant questions rather than absorbing the factual content at face value (without a sense of valid application). Students must assess each question and determine an optimal strategy to solve it, forming linkages between classes of questions and matching strategies, rather than repetitively applying a single rule to a large group of problems of a single type. Though this typically requires more effort and doesn't deliver the rapid improvement associated with massed practice of a single problem-solving strategy, it leads to better mastery and a increased retention of concepts for a longer time (\hyperlink{Roediger2013}{Roediger, 2013}). Recent studies (cf., \hyperlink{Rohrer2014}{Rohrer, Dederick, \& Burgess, 2014}) have found that interleaved rather than blocked practice can be almost twice as effective, highlighting the utility of this technique. 

Second, the library stimulates retrieval practice (\hyperlink{Bjork1975}{Bjork, 1975}; \hyperlink{Roediger2006}{Roediger \& Karpicke 2006}; \hyperlink{Soderstrom2013}{Soderstrom \& Bjork, 2013}) in its users through its basic format, a semester-long series of short review quizzes that constantly drive students to draw knowledge from memory and solve a variety of questions. Effective retrieval practice occurs when students employ it over a series of staggered sessions (rather than in single bursts of mindless repetition; \hyperlink{Roediger2013}{Roediger, 2013}). Our students may access a variety of aids to solve each question, as detailed above, but the knowledge that the formal weekly quizzes and the course exams will be conducted without these aids provides motivation for them to use aids initially, but over time strive to answer questions by drawing on memories instead, thus consolidating those memories. Because students can achieve 100\% scores for their review (homework) activities simply by meeting the target goal for number of problems solved (100 per week), these short quizzes can be treated as learning opportunities rather than a testing environment with pressure to meet a set standard score (particularly helpful for classes with a broad range of previous mathematical and scientific exposure). 

A second question of merit is how online tutors can help us to provide traditional science coursework to distance learners, to expand the reach of higher education and improve retention for dispersed populations. In a separate article (\hyperlink{Vogt2013}{Vogt, Cook, \& Muise, 2013}) we address one major challenge for the physical sciences, namely providing valid experimental laboratory experiences to students working remotely with no access to specialized laboratory equipment and training. We now present  evidence that our distance learners are performing at least as well as parallel in-class cohorts of students on lecture materials, with the aid of our GEAS online tutor.

Figure 3 shows the correlation between study within the self-review library and traditional exam performance for a remote cohort of distance learners and an in-class set of peers at NMSU. Exam scores are plotted against average review quiz (homework) scores at the time of each exam, showing a linear relationship in each case. 

\begin{figure*} [htbp]
  \begin{center}\epsfig{file=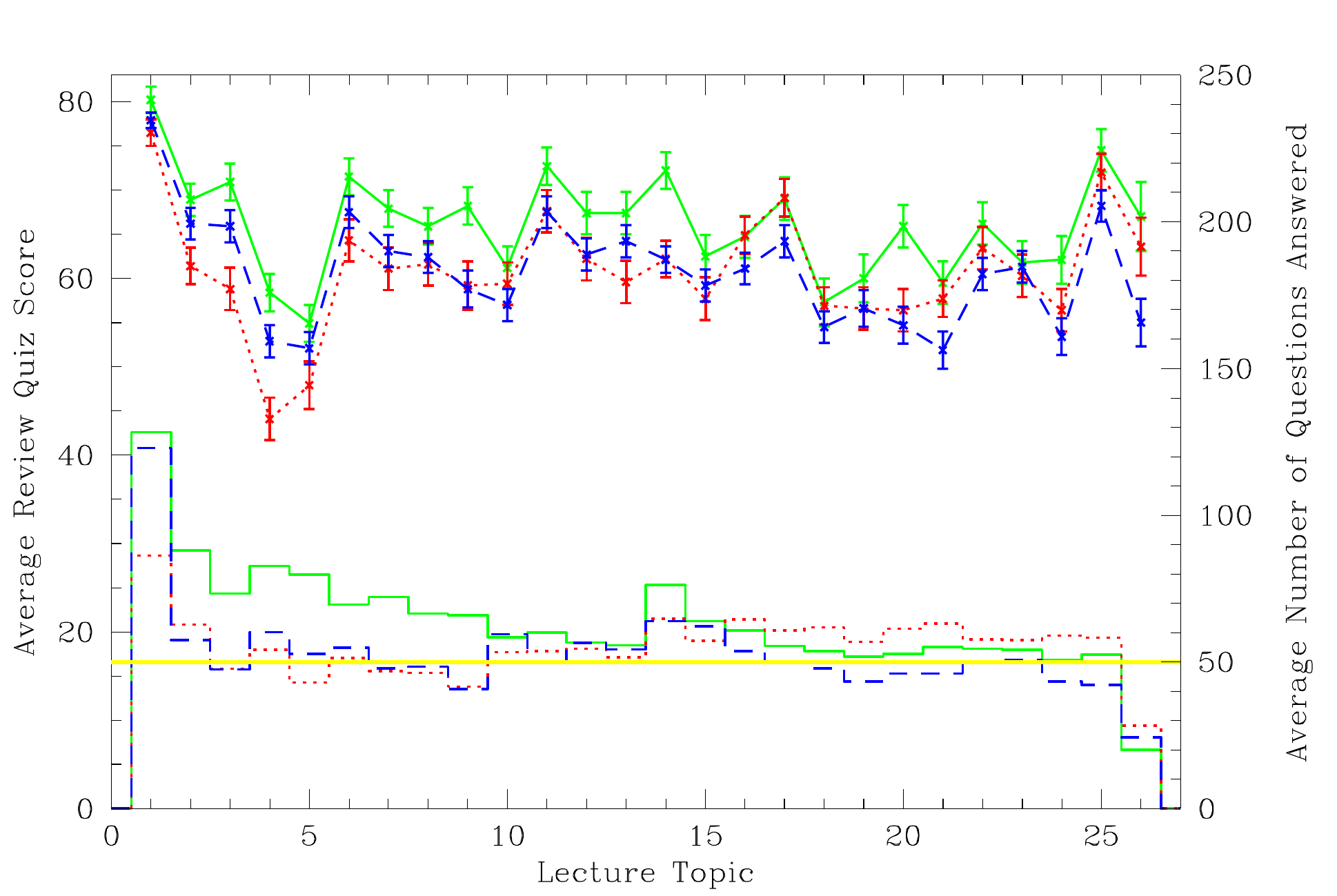, width=5.0in}\end{center}
  \caption[Figure 4] 
{Distribution of average scores (with 3 sigma error bars on each mean value) and number of questions attempted within the self-review library as a function of lecture module, for three student cohorts (distance learners in solid green, and two in-class cohorts in dashed blue and dotted red). The left axis labels average scores (line plots above), and the right axis labels the number of questions (histograms below). The horizontal yellow line represents the nominal goal of attempting 50 questions per lecture module and 1600 questions per semester. The peaks in number of questions attempted appear at the beginning of the course and the first lecture of the second half (after the midterm exam), while the low for the final lecture module reflects students shifting from weekly module study into second half review for the final exam. The pattern of high and low scores is reproduced in both cohorts, and is caused by variations in the difficulty of material from lecture to lecture. The distance learners attempt 10 more questions per lecture on average than the in-class cohorts; all groups hover around the nominal goal of 50 questions per lecture. They also score 5\% higher on average, in part due to completing more questions per lecture.}
  \label{fig04}
\end{figure*}
 
Figure 4 focuses on performance within the library for these two cohorts and another in-class group of undergraduates from Humboldt State University (HSU). We note that the three cohorts exhibit the same relative peaks and dips in average quiz score for various lectures, with the distance learners offset by +5\% relative to the in-class groups on average. They also complete 20\% more questions (averaging around 65 per lecture, with a nominal target of 50 for all groups), though note that the bulk of their extra studying takes place over the first half of the semester. 

\begin{figure*} [htbp]
  \begin{center}\epsfig{file=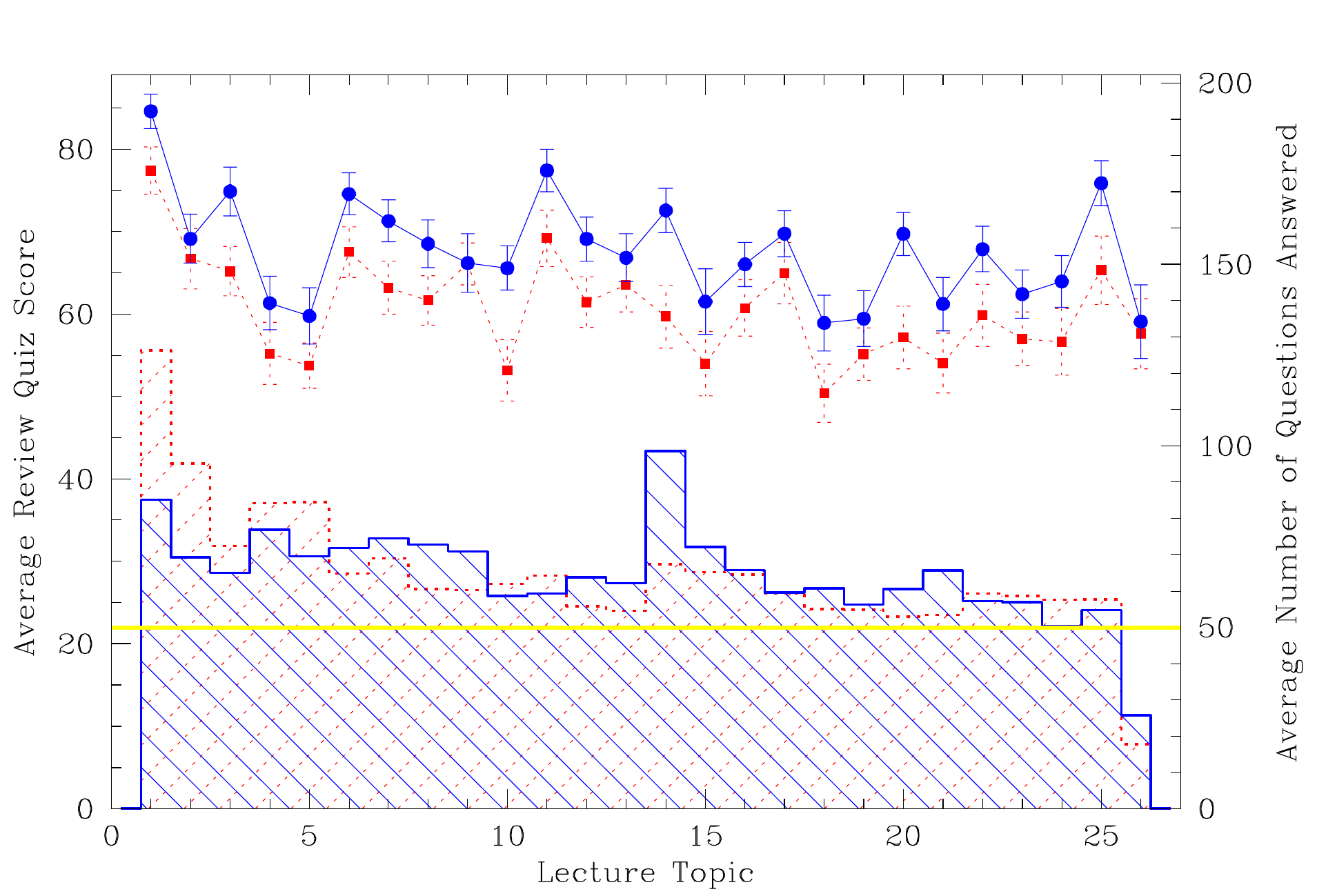, width=5.0in}\end{center}
  \caption[Figure 5] 
{Distribution of average scores and number of questions attempted within the self-review library as a function of lecture module, for distance learners. Figure format is otherwise as in Figure 4. Data are separated into results for men (blue solid lines) and women (red dotted lines). As in Figure 4, the pattern of high and low scores is reproduced in both cohorts, and is caused by variations in the difficulty of material from lecture to lecture. Male and female students complete similar numbers of questions (though women spend 8\% more time on them, for 61 minutes per week), but men score 7\% higher on average. }
  \label{fig05}
\end{figure*}

In Figure 5 we divide the distance learners into two groups by gender, and observe that within this cohort men outscore women by 7 $\pm$ 0.6\%  on average across the board. We do not, however, observe this offset in either of the in-class cohorts under study. 

The study of gender disparities in the physical sciences is a complex field unto itself. In a recent meta-analysis of 26 studies of concept inventories in physics, \hyperlink{Madsen2013}{Madsen, McKagan, and Sayre (2013)} grappled with the effects of 30 factors, including teaching and engagement techniques, student background and preparation, and stereotype bias. They found that no single factor could explain the differences in outcomes with gender. While the complete underlying causes of gendered effects are beyond the statistics available to us for our data set we can explain the observed difference between distance learners and classroom students, and suggest a possible cause for it as well.

Like Madsen, McKagan, and Sayre, we find an offset between men's and women's scores on a pre-class concept inventory. The difference is 16 $\pm$ 3\% for the distant cohort, but only half as much (7.9 $\pm$ 3\%) for the classroom students. The amount of time the genders spend studying within the self-review also differs between cohorts. For distant learners, men spend 5 minutes less per week answering questions (56 versus 61 minutes) but answer the same number of questions (65 per week, with a nominal goal of 50 per week). For the classroom students, however, men spend 11 minutes less per week answering questions (40 versus 51 minutes) and also answer significantly fewer questions (44 versus 62 per week). This suggests that weaker study habits of the male cohort in the classroom cohorts throughout the semester act to diminish the performance difference between genders. 

We speculate that one reason for this difference in observed study habits is a selection bias in course population. Our in-class cohorts contain 75\% freshman and sophomores and 25\% juniors and seniors, while these fractions are completely reversed in the distant cohort (which also tends to be a few years older). The more senior students in the distant cohort have less flexibility in fulfilling their general educational science coursework requirement (our astronomy course is currently the only qualifying science course at NMSU which can be completed remotely, while there are tens of options on-campus within astronomy, biology, chemistry, geology, and physics), and their clock are ticking on time to degree completion. They thus prioritize our course by necessity, as the cost of failure is higher.  

We can also address the issue of how student performance improves over time spent studying within the self-review library (see Figure 6). We analyzed the average score on questions over time within each lecture for each student in various cohorts, sorting questions in the order in which they were attempted and aligning the first question on each lecture for each student together, followed by the second question on each lecture for each student, and so on. These 1,600 curves of growth (for 26 lectures from 61 students) were then averaged together to create averaged curves showing the change in accuracy over time. Curves were truncated at the point at which only 30 individual samples remained to be averaged together, and also smoothed with a simple boxcar filter to remove the small-scale variations and highlight global behaviors. 

\begin{figure*} [htbp]
  \begin{center}\epsfig{file=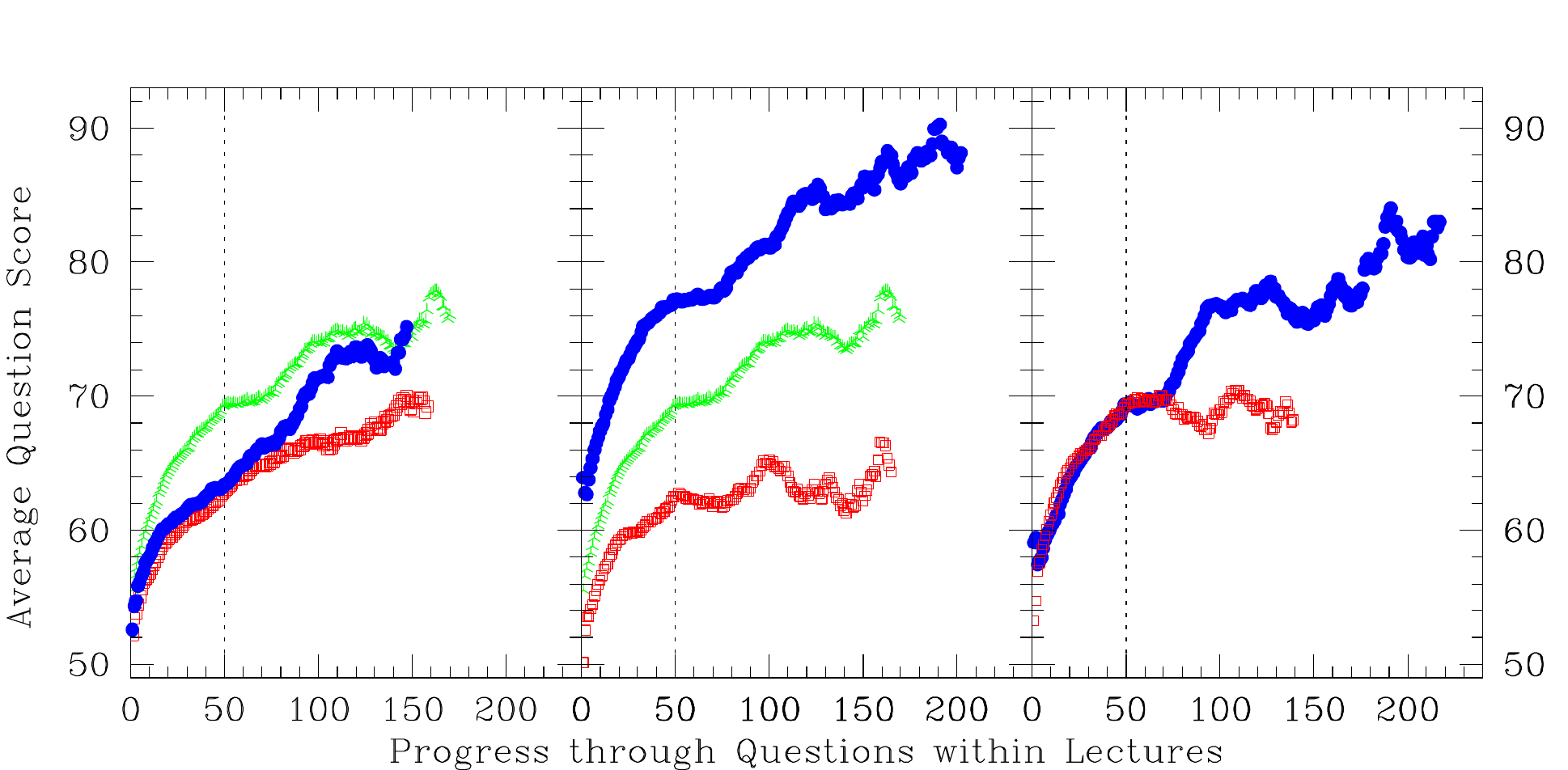, width=5.0in}\end{center}
  \caption[Figure 6] 
{Improvement in scores for review quiz questions (homework) over time, as more questions are answered within a given lecture module. Students start out with average accuracy levels around 55\%, increasing to 70\% or 80\% as they master new concepts and techniques for problem solving. In the left panel the green line of starred triangles indicates the progress for distance learners, while the blue solid dots and red open squares show results for two in-class cohorts. The distance learners start out at a slightly higher level of accuracy, and are performing significantly better by the 50-question mark (the nominal goal per lecture module, shown as a vertical line). For those students who continue to study, additional exposure results in significantly higher scores over time. Distance learners answer more questions on average, and end up with higher scores. In the middle panel, the cohort  of distance learners (green line of starred triangles again) starts out with accuracy levels around 55\%, increasing to almost 80\% on average as students master new concepts and techniques for problem solving. The cohort is then divided into two equal-sized groups based on final course grade. The high-grade group (blue dots) begins their studies with scores 12\% higher than the low-grade group (red open squares) and also makes considerably more progress over time, with scores 15\% higher at the 50-question mark and 20\% higher when the low-grade group stops studying, increasing by another 3\% as they complete an additional 40 questions on average. Students who perform better across the board in the class answer more questions on average, and end up with higher scores. In the right panel the distance learning cohort is redivided into students who complete the nominal goal of 320 quizzes (1600 questions) over a semester (red open squares) and those who complete more than 340 quizzes (blue solid dots).  The two curves match through the recommended 50-question period, but after this point the normal group plateaus in score and halts studying, while those who persevere longer more than double their improvement in score.}
  \label{fig06}
\end{figure*}

Our first, over-riding point is that the curves all rise. Over time, students learn new concepts and absorb new techniques for solving problems, and are able to apply them successfully. We have already shown that average scores within the library correlate well with traditional measures of performance (see Figure 3), and we can also observe the systematic acquisition of knowledge and skills over time within the learning environment. 

At first glance, one may notice that initial scores may begin as low as 55\% for some cohorts. This is due in part to the fact that these classes are for non-science majors, and host many students whose last contact with mathematics is a long-distant, faded memory. More important, however, is the way that we encourage students to structure their studying. Many of our students cling to traditional study habits (e.g., re-reading textbooks and underlining or highlighting key sections) even though the current understanding (\hyperlink{Roediger2013}{Roediger, 2013}) suggests strongly that these techniques are neither efficient nor beneficial to learning or performance.  We want our students to learn by doing (by combining diverse facts to draw conclusions, and by solving problems), and to utilize methods such as retrieval practice, a focus on interleaved topics, and pre-testing on new material to optimize their learning (cf., \hyperlink{Little2011}{Little \& Bjork, 2011}) as we best understand how to do so. We thus encourage them to begin their study of a new lecture by initially requesting a review quiz (beginning their homework) as a first step. Their initial scores can be quite low because of this, as they acquire information question by question, but the advantage is that they immediately see the applications and the purpose of facts under study. 

In the left panel of Figure 6 we present three curves for the cohorts shown in Figures 3--4, with distance learners in green and two in-class cohorts in red and blue (with colors as in Figure 4). For all three cohorts of students there is a clear increase in performance with time, and studying continues well past the nominal goal of 50 questions per lecture for each student. The distance learners complete more questions per lecture on average. Their average scores continue to rise with the additional exposure to material, but are also higher on average when they are compared to the in-class students at the same point along the curves (having answered the same number of questions on average). 

The middle panel of Figure 6 reproduces the same green curve showing the cohort of distance learners, and then splits this group into those with final grades in the top (blue) and bottom (red) halves of the group. We observe that all students improve performance over time, but those who perform best in the course begin at a higher level and also improve at a faster rate over the time interval beyond the 50-question limit (and continue to study additional questions, productively).

\begin{table} [htbp]
  \caption{Comparison of On-Campus and Remote Astronomy Cohorts}
  \begin{center}
  \begin{tabular} {c c c c} 
  \hline
  \hline
  Cohort & Exam Scores & Enrollment & Withdrawals \\
  \hline 
  Classroom & 76.6 $\pm$ 1.5 & 85 & 14 \\ 
  Remote      & 79.6 $\pm$ 1.7 & 97 & 36 \\ 
  \hline \\ 
  \end{tabular}
  \end{center}
  \label{tab02}
\end{table}

 The right panel of Figure 6 shows the same cohort of distance learners, again split into two groups. In this case we identified a group of ``super-studiers'', composed of 29\% of the students, those who completed between 340 and 754 quizzes over a semester (between 20 and 434, or 6\% and 136\%, beyond the nominal goal of 320). Because our grading algorithm guarantees students a score of 100\% for homework as long as students complete 20 quizzes per week (320 per semester), there is no direct increase in scores for any review work done beyond this limit. The students who choose to complete more review work thus appear to be doing so because they find it to be a beneficial strategy for studying. The two groups show identical improvement in accuracy up to the nominal 50-question goal line. Beyond this line, however, we observe that the normal group continue to study but shows no strong increase in accuracy, while the super-studiers answer twice as many questions beyond the nominal goal and continue to increase in accuracy over this interval. These students are not necessarily the strongest learners in the class nor those with the best backgrounds in math and science upon entering the course. They simply find this mode of study to work exceptionally well for them. 

Two versions of the NMSU general astronomy course were taught by the same instructor, one on-campus during Fall 2010 and one via distance learning from Fall 2011 -- Spring 2014. Written midterms and finals are never returned, so 88\% of the two-hour final exam was identical across both cohorts (see Table 2 and Figure 3). Distance learners performed slightly better, with a mean value two pooled standard deviations of the mean above that for the in-class students. (A two-sample t-test yields an inconclusive less than 9\% probability that the two scores were drawn from the same parent sample.) Our results are promising, and suggest that our distance learners perform at least as well as our in-class students.

Note that the distance learning cohort had a larger fractional rate of withdrawal (twice that of the in-class course), as expected (\hyperlink{Rovai2003}{Rovai, 2003}; \hyperlink{Woodley2004}{Woodley, 2004}). All students performed similarly on a pre-class astronomy concept inventory (\hyperlink{Halloun1985}{Halloun \& Hestenes, 1985}), and reported equal experience with computers. Those who completed the distance learning course were more likely to have studied math beyond algebra (though their self-assessed math skills were no higher). However, half of the students who withdrew did so without completing any course work, mainly due to difficulties securing financial aid. Other common reasons cited were difficulties spending enough time on the course, transfer to an in-class course, and a birth or death in the family. This is consistent with findings in the literature that non-academic factors dominate academic factors in driving withdrawal rates in distance learning (cf., \hyperlink{Subotzky2011}{Subotzky \& Prinsloo, 2011}).

\section{EXAMPLES OF SELF-REVIEW LIBRARY USAGE MODES}

There are four primary modes in which the self-review library is designed to be used. 

\subsection{Semester-Long Class Sequence}
The first mode applies to a class of students working through a full semester timeline. Students use the review and the weekly quiz tools and study all 26 lecture modules. This model works well for early-career faculty who have not yet invested significant time in developing their own lecture sequences, who opt to teach entirely from our materials. 

\vfill\eject
\subsection{General Study Tool}
The second mode applies to a class of students working on an independent timeline, covering many but not all of the 26 GEAS modules. These students do not take formal  weekly quizzes. They select sequences of modules to study based on their instructor's preferences (presented to them as default options at the correct times through a dedicated portal interface), and utilize the library as a personal study tool and to gain extra experience in working problems. This model works well for faculty with existing teaching materials of their own who wish to provide addition problem-solving experience to their students and/or to satisfy state or college requirements to provide external activities for students. 

\subsection{Targeted Activity}
The third mode applies to groups conducting a focused study of particular topics within the self-review library. These individuals work within a ``closed box'' subset of the library. All quizzes are conducted in review mode, and the progress reports and feedback forms are modified to reflect the short-term nature of the exercise. This model works well for studies evaluating new learning tutorials or aids. 

\subsection{Infrastructure Adaptation}
The fourth mode applies to instructors and developers interested in utilizing the framework of the self-review library to develop online tutorial offerings for other topics. The framework and interface controls of the library are cloned, and a parallel, independent archive is used to populate a self-review library for another field of study.

\vspace{0.20truein}
Collaborators are currently being solicited in all four modes, and inquiries are welcome from instructors interested in working with our materials in the classroom and for distance education cohorts. Individual test accounts in the self-review library are also available by request. More information on the General Education Astronomy Source (GEAS) may be found at http://astronomy.nmsu.edu/geas, or by contacting geas@astronomy.nmsu.edu.

\acknowledgments

This material is based in part upon work supported by the NSF through Grant No. AST-0349155 to NPV and by NASA through Grant No. NNX09AV36G to NPV. Any opinions, findings and conclusions or recommendations expressed in this material are those of the authors and do not necessarily reflect the views of the NSF or NASA.

We are pleased to thank the students and instructors of New Mexico State University and Humboldt State University who participated in our pilot program.

\section{Appendix}

Three attached figures show a sample self-review quiz (Figure 7), created in dynamic HTML by request, hint (Figure 8), and solution set (Figure 9). All such materials are archived within the class records for easy access by the instructor.


\begin{figure*} [htbp]
  \begin{center}\epsfig{file=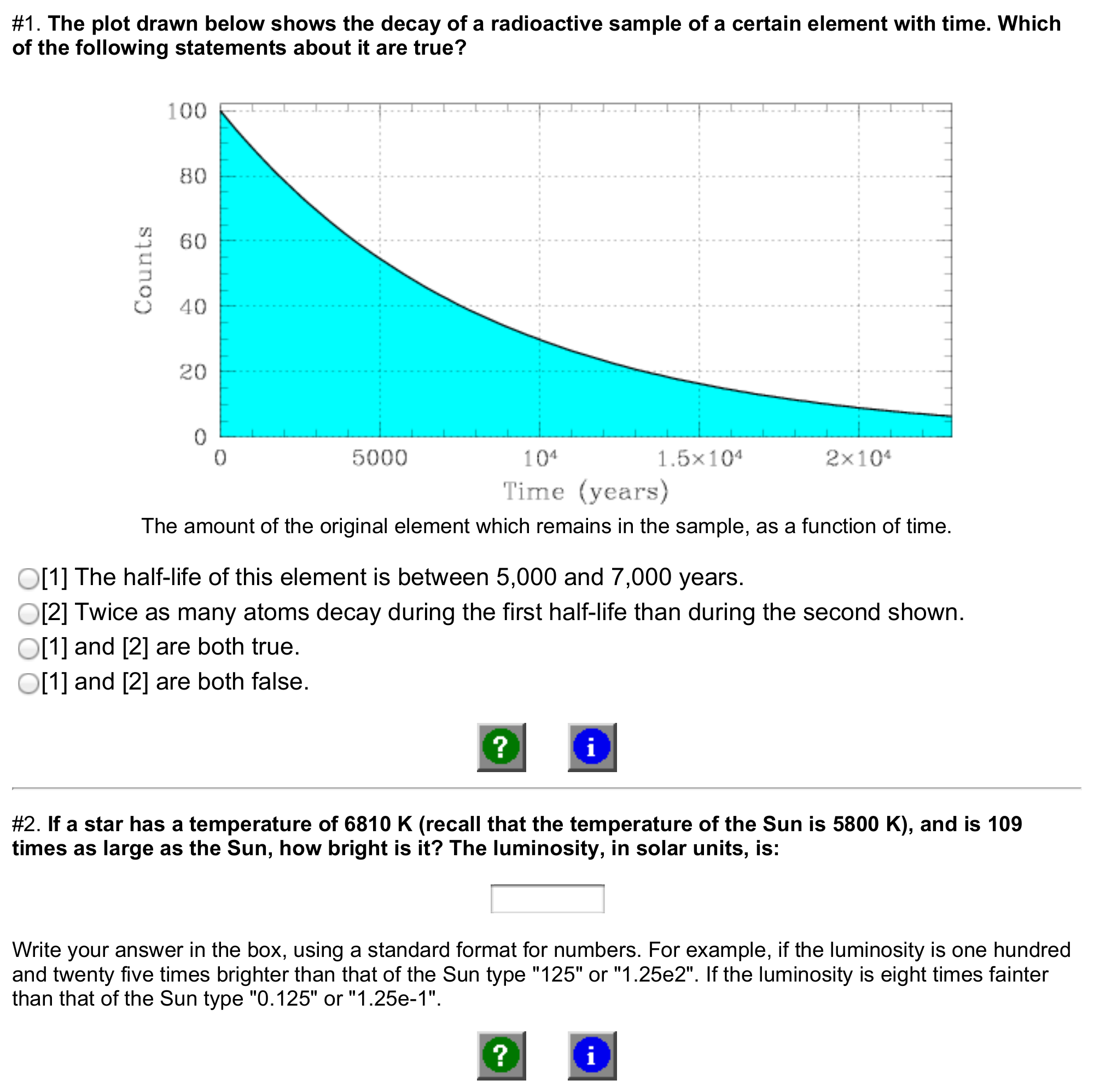, width=6.6in, clip=true}\end{center}
  \caption[Figure 7] 
{Sample questions from a self-review quiz, containing links to hints for solving the question (``?'' button), and a link back to the most relevant lecture slide (``i'' button). Both a representative multiple choice and numerical value question are shown. Radioactive decay is introduced as an age-dating mechanism as we discuss the formation of the solar system in Module 13, and the evolution of stars within the parameter space defined by luminosity (energy output), temperature, and size is the focus of Module 20.}
  \label{fig07}
\end{figure*}

\begin{figure*} [htbp]
  \begin{center}\epsfig{file=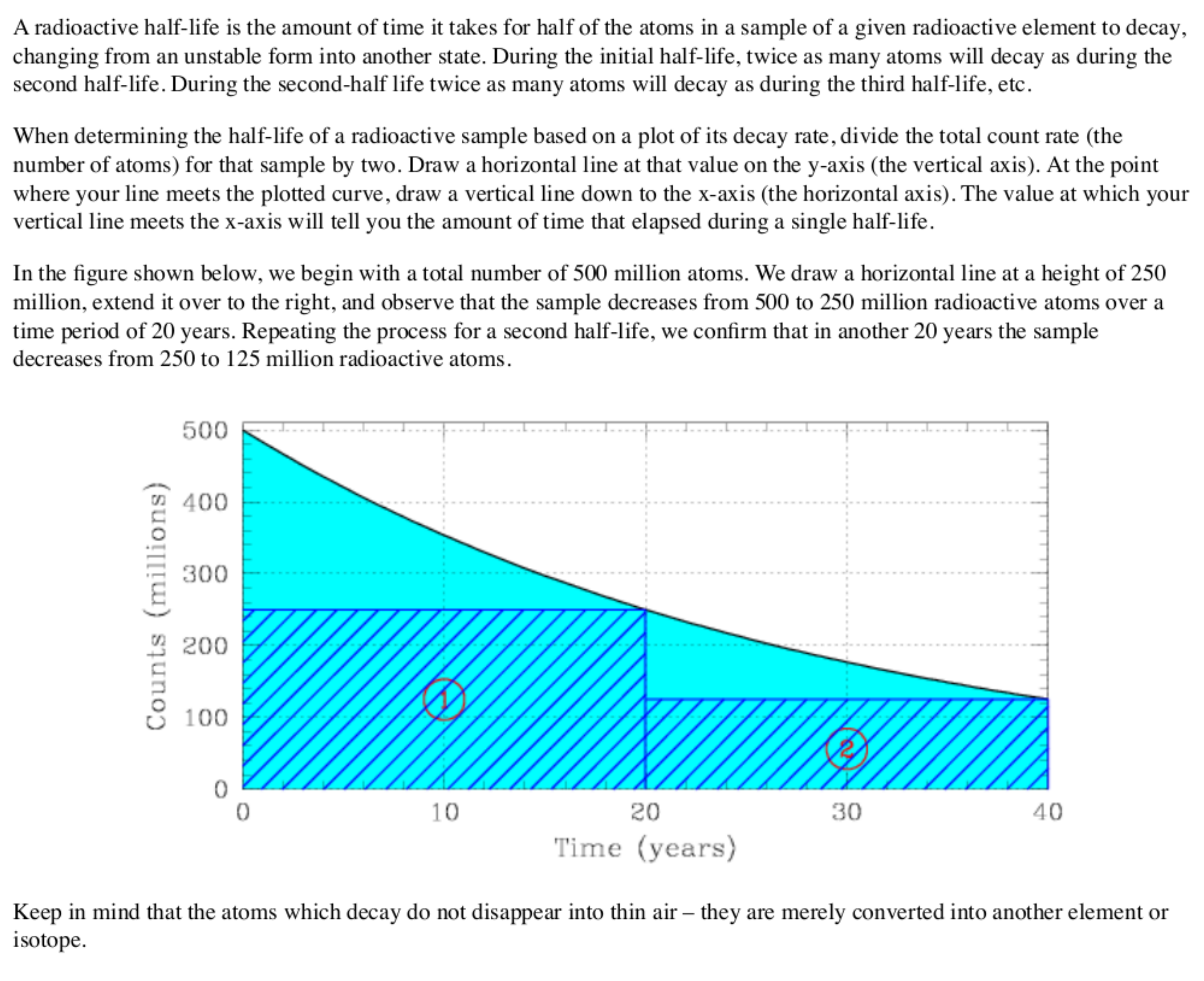, width=6.6in, clip=true}\end{center}
  \caption[Figure 8] 
{The hint for question \#1 shown in Figure 7, accessed by clicking on the ``?'' button within the question. Most students make frequent use of the hint option while studying, as well as the parallel link to the relevant lecture slide for each question.}
  \label{fig08}
\end{figure*}

\begin{figure*} [htbp]
  \begin{center}\epsfig{file=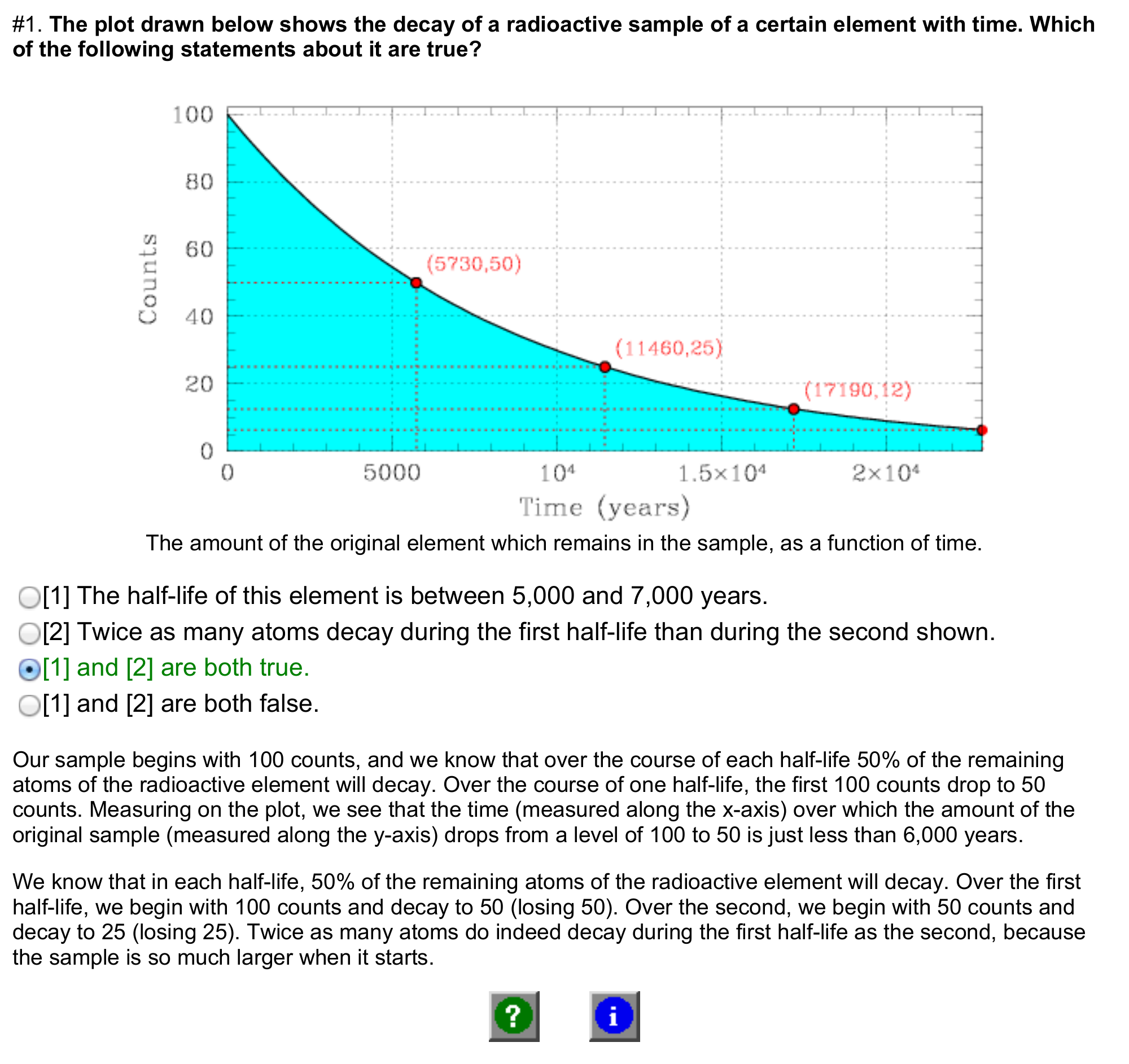, width=4.8in}\end{center}
  \begin{center}\epsfig{file=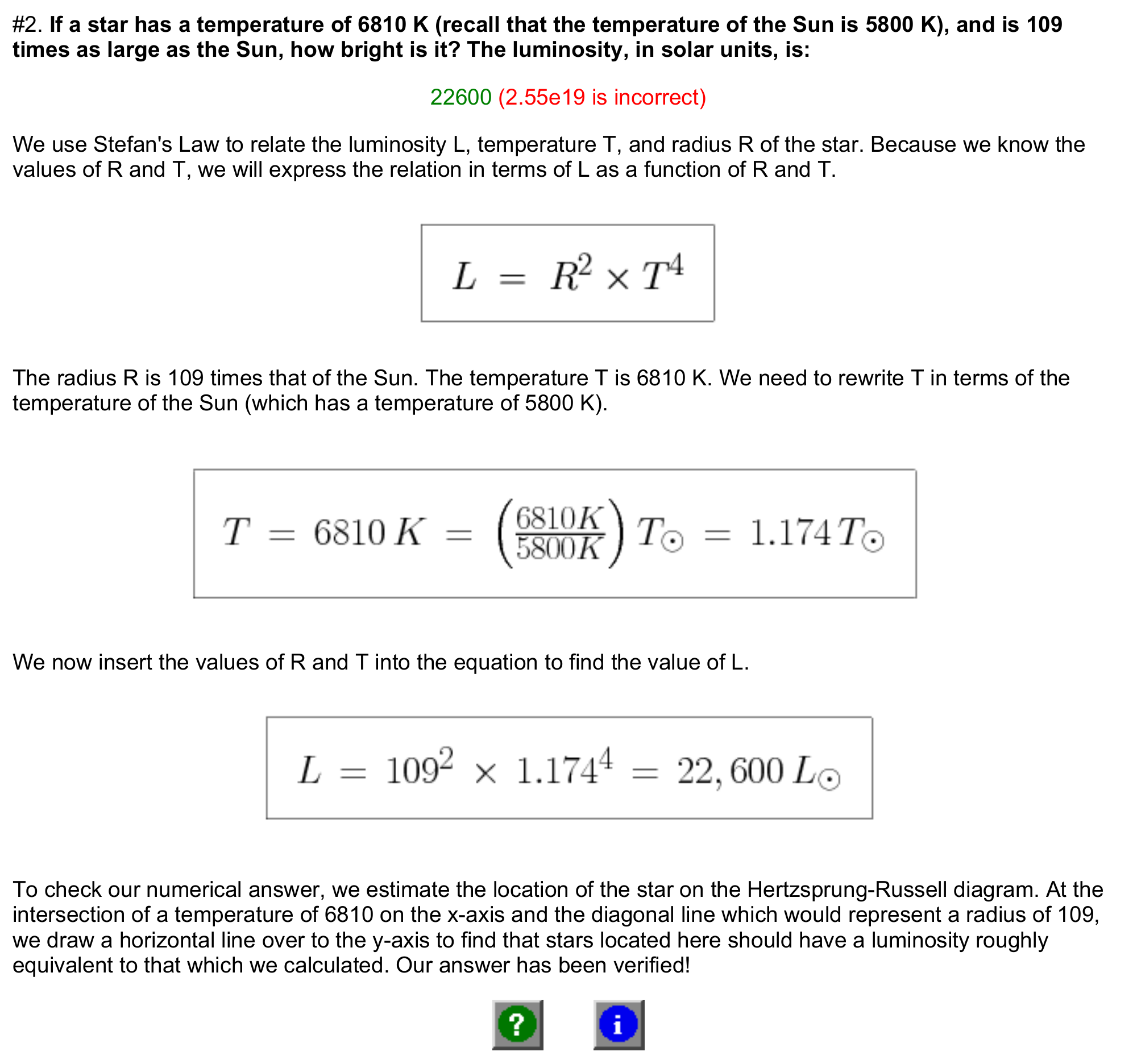, width=4.8in}\end{center}
  \caption[Figure 9] 
{A portion of a solution set, for the two quiz questions shown in Figure 7. A worked solution is presented for each problem, noting both student answers and correct ones. The figure in question \#1 has been annotated to help students to understand it; figure reading is a new skill for many of our students. In question \#2, the student neglected to convert the stellar temperature from kelvins into solar units, resulting in an absurdly large luminosity. The value needed to lie within 250 $L_{\odot}$ of 22,600 $L_{\odot}$ in order to be accepted as correct, giving a reasonable amount of room for rounding errors in the calculation.}
  \label{fig09}
\end{figure*}

\end{document}